\documentstyle[prl,multicol,epsf,epsfig,aps]{revtex}
\newcommand{\BEQ}{\begin{equation}}
\newcommand{\EEQ}{\end{equation}}
\newcommand{\BEA}{\begin{eqnarray}}
\newcommand{\EEA}{\end{eqnarray}}

\renewcommand{\d}{{\rm d}}

\newcommand{\g}{{g}}

\newcommand{\si}{\hat{\sigma}}
\newcommand{\om}{\omega}
\newcommand{\alp}{i}
\newcommand{\bet}{j}
\newcommand{\ab}{{\alp\bet}}

\newcommand{\tr}{{\rm tr}}

\newcommand{\HPO}{\hat{A}}
\newcommand{\PO}{A}
\newcommand{\HOB}{\hat{s}}
\newcommand{\OB}{s}

\newcommand{\half}{\frac{1}{2}}

\newcommand{\CD}{{\cal D}}

\newcommand{\B}{\hat{B}}
\newcommand{\La}{\Lambda}

\newcommand{\hm}{\hat{m}}

\newcommand{\su}{\hat{s}}
\newcommand{\HH}{\hat{H}} 
 
\newcommand{\CR}{{\cal R}}
\newcommand{\RS}{{\rm S}}
\newcommand{\RA}{{\rm A}}
\newcommand{\RM}{{\rm M}}
\newcommand{\RB}{{\rm B}} 
\newcommand{\ri}{{\rm i}} 
\newcommand{\down}{{\downarrow}}
\newcommand{\up}{{\uparrow}}
\newcommand{\uu}{{\uparrow\uparrow}}

\newcommand{\ud}{{\uparrow\downarrow}}
\begin{document} 
\draft
\title{Curie-Weiss model of the quantum measurement process} 
\author{Armen E. Allahverdyan$^{1,2)}$, Roger Balian$^{1)}$ 
and Theo M. Nieuwenhuizen$^{3)}$}

\address{$^{1)}$ SPhT, CEA Saclay, 91191 Gif-sur-Yvette cedex, France
\\ $^{2)}$Yerevan Physics Institute,
Alikhanian Brothers St. 2, Yerevan 375036, Armenia\\
$^{3)}$ Institute for Theoretical Physics, % University of Amsterdam\\
Valckenierstraat 65, 1018 XE Amsterdam, The Netherlands} 

\maketitle

\begin{abstract}
A hamiltonian model is solved, which satisfies all  requirements
for a realistic ideal quantum measurement. The system $\RS$ is a spin-$\half$,
whose $z$-component is measured through coupling with an apparatus 
$\RA=\RM+\RB$, consisting of a magnet $\RM$ formed by a set of
$N\gg 1$ spins with quartic infinite-range
Ising interactions,  and a phonon bath $\RB$ at temperature $T$.
Initially $\RA$ is in a metastable paramagnetic phase.
The process involves several time-scales.
Without being much affected, $\RA$ first acts on $\RS$, whose state
collapses in a very brief time. The mechanism differs from the usual 
decoherence. Soon after its irreversibility is achieved. Finally
the field induced by $\RS$ on $\RM$,
which may take two opposite values with probabilities given by Born's rule, 
drives $\RA$ into its up or down ferromagnetic phase.
The overall final state involves the expected correlations between the
result registered in $\RM$ and the state of $\RS$.
The measurement 
is thus accounted for by standard quantum statistical 
mechanics and its specific features arise from the macroscopic size
of the apparatus.

\end{abstract}


\begin{multicols}{2}

The quantum measurement problem has given rise to an immense 
literature \cite{wh}, but it is still an object of debate. Insight 
can be gained by studying exactly solvable 
dynamical models \cite{models}. The one
we consider hereafter is intended to be as realistic as possible. 

We first have to face the irreducibly probabilistic
nature of quantum mechanics, arising from the non-commutation of 
observables. Within the statistical interpretation, 
see e.g.  \cite{ballentine},
a wave-function can provide us with nothing more than the probabilities 
for the values of any physical quantity. We adhere to the so-called 
bayesian view on probabilities \cite{bayes}: they are not inherent to a single 
object but refer to a population to which it belongs
(in reality or in thought); they are
mathematical tools for deducing sensible predictions from given prior
knowledge. Thus, what is called a quantum ``state'', whether pure or mixed,
gathers our information on the considered system: it does not
characterize this system by itself, but its statistical ensemble.
Accordingly, the consistency of
quantum mechanics requires a measurement to be analyzed as a statistical
process involving many similar experiments with all possible outcomes. 
This process couples a system (S) to an apparatus (A),
creating correlations between the initial state (in the above sense) of S
and the final value $A_i$ of a pointer variable $\hat{A}$ of A. One infers
thereby information about S as regards the occurrence of the corresponding
eigenvalue $s_i$ of one of its observables
$\hat{s}$. This knowledge has statistical nature because quantum
theory cannot inform us fully about a single object, even if it is 
``completely'' prepared in a pure state. 

Quantum statistical mechanics is also unavoidable for other reasons. 
On the one hand, the apparatus should be {\it macroscopic} so as to 
ensure registration of the outcomes $A_i$. The recorded results may 
later on be read by an observer whom we thus can leave aside. 
The required large size of A forbids us to assign a pure state to it
since it cannot be completely prepared at the microscopic level.
Rather, as always in statistical problems, 
{\it we must assign to it a density matrix}. 
On the other hand, a quantum measurement is an {\it irreversible} process. 
Statistical mechanics is necessary to explain this specific type of 
irreversibility, as any other one, by relying on microscopic {\it hamiltonian} 
dynamics. 

Even if the coupling of S and A is weak, the perturbation
it induces on S cannot in general be neglected in a quantum measurement. We 
will consider a so-called ideal measurement, which perturbs S as little as 
possible, keeping unchanged the statistics of all observables which commute 
with $\HOB$. The final density matrix of ${\rm S}$, $r(t_{\rm f})$, 
is then obtained from $r(0)$ by cancelling
the off-diagonal blocks associated with different eigenvalues
$\OB_i$ and $\OB_j$ of $\HOB$. We denote the remaining diagonal blocks 
as $r_i$.

To represent a measurement, a process should have several specific features
\cite{wh,balian,vKampen}.
{\it i}) The apparatus $\RA$ is macroscopic and at the initial time $t=0$ 
in a metastable state $\CR(0)$, independent of the arbitrary state $r(0)$ 
of the system $\RS$. The full density operator has the form
\BEA
\label{CalD0=}
\CD(0)=r(0)\otimes \CR(0).
\EEA
{\it ii}) Triggered by its coupling
with $\RS$, $\RA$ may reach at the end of the process one among
several possible states $\CR_i$, which are a priori equally probable so as to
avoid any bias. {\it iii}) Each state $\CR_i$ is stable so as to register 
information robustly and permanently. The pointer variable $\HPO$ has in the
state $\CR_i$ negligible fluctuations around $\PO_i$ 
so as to ensure precise and clear distinction between the possible outcomes. 
{\it iv}) The observable $\hat s$ of $\RS$ does not change much during 
the process, and thus nearly commutes with the Hamiltonian. 
{\it v})  An ideal measurement involves a collapse, 
which changes $r(0)$ into the sum of its diagonal blocks $r_i$ 
corresponding to each $\OB_i$. 
{\it vi}) The density operator at the end of the process
involves a special type of classical 
correlations between $\RS$ and $\RA$, namely, 
\BEA \label{D0toDinf}
\label{2}
\CD(0) \mapsto \CD(t_{\rm f})=\sum_ir_i\otimes \CR_i
\equiv \sum_ip_i\times\frac{r_i}{{\rm tr}\,r_i}\otimes \CR_i.
\EEA
Together with {\it iii} and with probabilistic 
interpretation of a quantum state, Eq.~(\ref{2}) means that
the pointer variable may take any of the values $A_i$ corresponding to
$\OB_i$ with the 
probabilities $p_i={\rm tr}\,r_i$ (Born's law), and that if we select
$A_i$, the subsequent statistics of $\RS$ is described by the density 
operator ${r_i}/{{\rm tr}\,r_i}$ (von Neumann's reduction). 

Most of these features have been emphasized in the past, 
and the models already worked out exhibit some among them
\cite{wh,models,ballentine,balian,vKampen}. In the present model 
{\it all} the above requirements will be fulfilled.
In order to satisfy conditions {\it ii}
and {\it iii} we take for $\RA$ a macroscopic system displaying
a phase transition with broken symmetry so as to eliminate any bias; 
$\HPO$ is an order parameter
with small fluctuations. Moreover, $\RA$ involves a 
large number of degrees of freedom which ensure 
an irreversible relaxation towards one of the equilibrium states $\CR_i$. 

{\it The model.}
Our system $\RS$ is a spin-$\half$, the observable to be measured 
is its third Pauli matrix $\HOB_z$ with eigenvalues $\OB_i$ equal to $\pm 1$.
Our apparatus $\RA=\RM+\RB$ simulates a magnetic dot: 
$\RM$ consists of $N\gg 1$ spins with Pauli operators $\si^{(n)}_a$ ($a=x,y,z$)
coupled to the phonon bath $\RB$. The order parameter $\HPO$ is the
magnetization in the $z$-direction $\hm=\frac{1}{N}\sum_{n=1}^N\si^{(n)}_{z}$.
The Hamiltonian $\HH=\HH_\RA+\HH_{\RS\RA}$
governing the measurement process reads:
\BEA
\label{hamham}
&&\HH_\RA=\HH_{\RM}+\HH_{\RM\RB}+\HH_{\RB},
\quad\HH_{\RS\RA}=-\g N\hm\su_z, \\
&& \HH_{\RM}=-\frac{1}{q}\,JN\hm^q, 
\quad\HH_{\RM\RB}=\sqrt{\gamma}\,\sum_{n=1}^N \sum_{a=x,y,z}
\si^{(n)}_{a}\B_{a}^{(n)}.\nonumber
\EEA
It commutes with $\hat s_z$ in agreement with
\cite{wigner}, ensuring the condition {\it iv} above. 
The interaction $\HH_{\RS\RA}$ is a spin-spin coupling with strength $g>0$.
The Hamiltonian $\HH_{\RM}$, of the 
Curie-Weiss-type, couples all spins symmetrically. 
For $q=2$ one has an Ising model, while 
$q=4$ or $6$ describes so-called super-exchange 
interactions as realized for metamagnets. We take $q=4$.
The bath $\RB$ describes a set of Debye phonons at equilibrium
in the thermodynamic limit. The dimensionless constant $\gamma$  
characterizes the strength of their coupling to the spins of ${\RM}$.
It should be weak to trace out B and to ensure an exact Boltzmann-Gibbs 
distribution of M at equilibrium.
The properties of the bath operators $\B_{a}^{(n)}$ will be specified later. 

{\it Mean-field approximation for the apparatus.}
The long-range character of $\HH_{\RM}$ ensures that its 
equilibrium behavior is exactly described in the large-$N$ limit by 
the mean-field approximation. However, the correlations between $\RS$ 
and $\RA$ induced by $\HH_{\RS\RA}$ are essential for our purposes. 
Their very existence prevents us from applying
the standard mean-field method. A way out of this 
is to separate the state $\CD$ of the total system $\RS+\RM+\RB$
into several sectors, and 
use a {\it different} time-dependent mean-field in each of them. 
In the eigenbasis of $\su_z$ for ${\rm S}$, 
$|i\rangle =|\up\rangle$ or $|\down\rangle$ for eigenvalues
$s_i=\pm 1$,
$\CD$  has the elements $\CD_{\alp\bet}=\langle \alp |\CD|\bet\rangle$. 
The von Neumann equation reads for each of the operators
$\CD_{\alp\bet}$ in the $\RM+\RB$ space
\BEA
\label{klein2}
\ri\hbar\frac{\d}{\d t}\CD_{\alp\bet}&=&
-\g N(s_\alp\hat{m}\CD_{\alp\bet}-s_\bet \CD_{\alp\bet}\hat{m})
+ [\HH_\RA,\CD_{\alp\bet}].
\label{klein3}
\EEA
The mean-field approach is implemented for each 
$\CD_\ab$. In $\HH_\RA$  we replace $\hm^4$ 
by $m_\ab^4+4m_\ab^3(\hm-m_\ab)$, where the $c$-number $m_\ab$ is 
determined selfconsistently. To do this we note that
(unlike $\CD_{\up\up}$ and $\CD_{\down\down}$)
$\CD_{\up\down}$ and $\CD_{\down\up}=\CD_{\up\down}^\dagger$ 
are neither positive nor hermitian. Taking 
\BEA
m_{\alp\bet}=\frac{ {\rm tr}\,\hat{m}\,|\CD_\ab|}
{ {\rm tr}\,|\CD_\ab|},\qquad |\CD_{\alp\bet}|\equiv
\sqrt{\CD_{\alp\bet}\CD^\dagger_{\alp\bet}}
\label{kaa}
\EEA
for any pair $\ab$, one can show that this approximation becomes exact
for large $N$, as in the static case.

{\it Properties of the bath.}
The interaction between the bath ${\RB}$ and the magnet ${\RM}$ 
is treated within the cumulant weak-coupling approach, see e.g. 
\cite{gardiner}. The initial state of ${\RB}$ 
is gibbsian at temperature $T=1/\beta$, $R_{\rm B}(0)
=\exp(-\beta \hat H_{\rm B})/Z_{\rm B}$, and is
not correlated with the density operator $D(0)=r(0)\otimes R_\RM(0)$
of $\RS+\RM$.
The Hamiltonian $\HH_{\RB}$ is such that the free correlation functions 
of the bath variables are stationary, identical and
independent for different components $a$ and different sites $n$, 

\BEA
{\rm tr}_{\rm B}[\,R_{\rm B}(0)\B_{a}^{(n)}(t)\B_{b}^{(m)}(s)\,]
=\delta_{a,b}\,\delta_{n,m}\,K(t-s).
\EEA
In agreement with the initial equilibrium of ${\rm B}$, we
assume a quasi-ohmic spectrum
\BEA
K(t)=\hbar^2\,\int_{-\infty}^\infty \frac{\d \om}{16\pi} \,e^{\ri \om t}\,
\om\,(\coth \half\beta\hbar\om-1)\,{\rm e}^{-|\om|/\Gamma},
\label{damask4}
\EEA
with $\Gamma$ the Debye frequency cut-off, where
$\hbar \Gamma$ exceeds all other energy scales: $T$, $J$ and $g$.
The correlation time of $\RB$, of order $\hbar/T$, is 
much smaller than the characteristic times of $\RA$
variables, which for long times ensures relaxation of $\RA$ 
towards the Gibbs distribution.
Besides the weak-coupling limit $\gamma\ll 1$, 
this requires not too low  a temperature $(T\gg \gamma J)$.

{\it Bloch equations.} As indicated above,  in the large $N$ limit
each block of $\CD$, and hence of $D=\tr_\RB\,\CD$, is
of the time-dependent mean-field type:
\BEA
\label{HF}
D_{\alp\bet}(t)=r_\ab(0) \times
\rho^{(1)}_{\alp\bet}(t)\otimes\cdots\otimes 
\rho^{(N)}_{\alp\bet}(t).
\EEA
Each $\rho^{(n)}_{\alp\bet}$ lives in the $2\times 2$ Hilbert space of 
the $n$'th spin of ${\RM}$, and reads in the polarization representation: 
$\rho_{\alp\bet}^{(n)} =
% \half\zeta_{0,\,\alp\bet}\,\si^{(n)}_0+
\half\sum_{a=0,x,y,z}\zeta_{a,\,\alp\bet}\,\si^{(n)}_a$, where
the $\si^{(n)}_0$ are $2\times 2$ identity matrices and
where $\zeta_{a,\,\alp\bet}={\rm tr}\,\si_a^{(n)}\rho^{(n)}_{\alp\bet}$ 
is independent of $n$.

As the coupling is weak,
we can eliminate the bath from (\ref{klein2}), 
even for times shorter than the memory-time of $K(t)$;
Eq.~(\ref{HF}) yields $\zeta_{x,\,ij}=\zeta_{y,\,ij}=0$ at all $t$
and
\BEA
&&\dot{\zeta}_{0,\ud}=\frac{2\ri \g}{\hbar}\zeta_{z,\ud},\quad
\dot{\zeta}_{z,\ud}=\frac{2\ri \g[1+\eta(t)]}{\hbar}\zeta_{0,\ud}
-2\La(t)\,\zeta_{z,\ud},\nonumber\\
&&
\label{birma9}
\EEA
where for $t\ll1/\Gamma$, $\La$ equals $\gamma\Gamma^2t/2\pi$ 
and $\eta=\gamma\Gamma^2t^2/2\pi$.
They go in the markovian limit $t\gg\hbar/T$ to
$\La(\infty)=\gamma g/(2\hbar\tanh\beta g)$ and $\eta(\infty)= 0$.

For $\zeta_{0,\uu}$ and for the magnetization $m_{\up}=\zeta_{z,\up\up}/
\zeta_{0,\uu}$ we shall need below the only markovian equation: 
\BEA
\dot{\zeta}_{0,\uu}=0,\quad
\dot{m}_{\up}=
\frac{\gamma h_{\up}}{\hbar}(1-\frac{m_{\up}}{\tanh\beta h_{\up}}),
\label{rangun}
\EEA
where $h_{\up}(t)=\g+Jm_{\up}^{3}(t)$ is the effective field 
in the considered sector.The sign of $g$ is changed
for the $\down\down$ sector.

{\it Initial conditions.}
Before the measurement, ${\RM}$ is prepared in a paramagnetic state 
$R_\RM(0)=2^{-N}\prod_n\si^{(n)}_0 $, leading to the
initial density matrix of $\RA=\RM+\RB$: $\CR(0)= R_\RM(0)\otimes R_\RB(0)$.
The measurement will be unbiased, 
since the order parameter vanishes initially,  $m_{\alp\bet}(0)=0$. 
According to Eqs. (\ref{CalD0=}, \ref{HF}) we 
have $\zeta_{0,\,\ab}(0)=1$, $\zeta_{a,\,\ab}(0)=0$.
The paramagnetic state is metastable for temperatures below 
 $0.36 J$ and this is the regime where A can act as a measuring apparatus.
 Its decay-time is then exponentially large in $N$, thus basically infinite.
However, a change in the macroscopic state of the $\RA$ can be induced 
by its coupling with $\RS$, provided $g$ is sufficiently large so as 
to suppress the barrier leading to the lowest ferromagnetic state.
% ($g>0.08J$ for  $T=0.34J$).  

{\it Collapse.} 
We first consider the evolution of the off-diagonal elements
$r_{\up\down}$, given by (\ref{HF}) as
\BEA
\label{rudt} r_{\up\down}(t)=
\tr_{\RM,\RB}\CD_{\up\down}(t)=\tr_{\RM}D_{\up\down}(t)=
r_{\up\down}(0)\,\zeta_{0,\up\down}^N(t).
\EEA
Obtained by solving (\ref{birma9})
with its initial conditions by the exact WKB-method,
$\zeta_{0,\,\up\down}(t)$ has the form 
$\exp[-\chi(t)]\cos\theta(t)$, where
$\theta(t)=\frac{2g}{\hbar}\int_0^t\d s\,\exp 2[\chi(s)-\int_0^s
\d u\,\Lambda(u)]$.
For $t<1/\Gamma$, $\chi$ and $\theta$ behave as
\BEA
\label{litovsk}
\chi\approx\frac{\gamma\Gamma^2g^2}{2\pi\hbar^2}\,t^4,\quad
\theta\approx\frac{2gt}{\hbar}\left(1-\frac{\gamma\Gamma^2t^2}{6\pi}
\right).
\EEA
(The markovian regime $t\gg \hbar/T$,  $\chi=\Lambda(\infty)t$ 
is irrelevant here when $N$ is very large.)
The amplitude of $\zeta_{0,\ud}$ decreases as an exponential,
quartic in $t$ for small times and linear for long times. 
Since $\zeta_{z,\ud}$ is imaginary, $\CD_{\up\down}\CD_{\up\down}^\dagger$
is proportional to the unit matrix, so that $m_{\up\down}$ vanishes.

We thus find for $r_{\up\down}(t)=r_{\up\down}(0){\rm e}^{-N\chi(t)}
[\cos \theta(t)]^N$ a sequence of narrow gaussian peaks,
arising from the nearly periodic cosine factor 
and located around the times at which $\theta/\pi$ is an integer.
At the very beginning of the measurement,
(\ref{litovsk}) show that the off-diagonal elements
$r_{\up\down}(t)=r_{\down\up}^*(t)$ of the marginal density matrix 
of S rapidly fall down.
For a large apparatus such that $N\gg \gamma\,(\hbar\Gamma/g)^2$, 
this decrease is dominated by that of $[\cos (2gt/\hbar)]^N
=\exp[-(t/\tau_{\rm collapse})^2]$ rather than that of $\exp[-N\chi]$.
The collapse time
\BEA
\label{taucollaps}
\tau_{\rm collapse}=\frac{1}{\sqrt{2N}}\,\frac{\hbar}{g},
\EEA
thus characterizes the disappearance of the components
$D_{\up\down}$ and $D_{\down\up}$ for S+M.
This time is much shorter than all other characteristic 
times of the process.
Its coefficient $\hbar/g$ differs from the one
$\hbar/T$ that enters standard decoherence times.
 
The subsequent spikes, the first of which occurs at $t=\pi\hbar/(2g)$,
are suppressed by the factor $\exp[-N\chi(t)]$ which arises due to
the interaction with the bath, a decay amplified by the large value of $N$.
Due to (\ref{litovsk}) this means that after a decoherence time
%$\tau_{\rm decoh}=(2\pi/\gamma N)^{1/4}(\hbar/\Gamma g)^{1/2}$
\BEQ \label{taudecoh}
\tau_{\rm decoh}=\left(\frac{2\pi}{\gamma N}\right)^{1/4}
\left(\frac{\hbar}{\Gamma g}\right)^{1/2}
\EEQ
all the peaks are washed out, and $r_{\up\down}$ and $r_{\down\up}$ remain 
zero after their initial gaussian collapse, while $r_{\up\up}$ and $r_{\down\down}$ 
are kept unchanged since $\hat{s}_z$ is conserved. The collapse proper thus results
only from the interaction $\hat{H}_{\rm SA}$ of S with the large 
number of spins, as shown by (\ref{taucollaps}). 
The irreversible loss of information about the off-diagonal elements takes place 
at a characteristic time $\tau_{\rm decoh}$, after the collapse. 
Provided  $N\gg (\hbar\Gamma/g)^2/\gamma$, it is faster than $1/\Gamma$,
the attempt time of the bath, and well before the recurrence 
of the first peak, since  we assumed already that $\hbar\Gamma\gg g$. 
This irreversibility is essential although hidden, since $r_{\up\down}$
has not yet revived when it takes place.

%after the collapse but before the first recurrence time.
% This irreversibility is essential although hidden, since $r_{\up\down}$
% has already become  negligible at the time when it occurs. 

In fact, not only the initial collapse, but even the  
suppression of possible revivals, do not require M to interact with a bath.
A realistic interaction $\hat{H}_{\rm SA}$ would involve a small dispersion
$\delta g$ around the average value $\langle g\rangle$ of the coupling 
between $\hat{s}_z$ and the various spins $\hat{\sigma}_z^{(n)}$ of A. 
For independent disorder a cumulant expansion now brings 
$\theta =\langle g\rangle 2t/\hbar-\langle\delta g^3\rangle_{\rm c} (2t/\hbar)^3/3!
+\cdots$ and 
$\chi=\langle\delta g^2\rangle_{\rm c} (2t/\hbar)^2/2-
\langle\delta g^4\rangle_{\rm c} (2t/\hbar)^4/4!+\cdots$.
The damping factor $\exp[-N\chi(t)]$ then suppresses the recurrent peaks 
provided $N\gg \langle g\rangle^2/\langle\delta g^2\rangle_{\rm c}.$
The large size of the apparatus thus suffices to make 
the initial collapse permanent. The irreversibility of the collapse is
here as a collective effect due to the large value of $N$ and not
to the environment. These two mechanisms  for suppression of 
the $x,\,y$ components of the spin S are reminiscent of the spin-lattice 
and spin-spin relaxation mechanisms in NMR, respectively.

{\it Registration.}
Let us now consider the evolution of the diagonal elements
$D_{\up\up}$ and $D_{\down\down}$. Since they are not affected by the
initial process, their evolution is governed by (\ref{rangun}) in the 
markovian regime $t\gg \hbar/T$. The registration by the apparatus will 
therefore look like a relaxation towards equilibrium in statistical mechanics. 
The Curie-Weiss  equation, $m_{i}=\tanh[\beta h_{i}]$ with
$h_i=\pm g+Jm^{3}_{i}$ for each sector $i=\up$ of $\down$,
 gives the extrema of the free energy per site
$F_i(m)=\mp gm-\frac{1}{4}Jm^4-TS(m)$, where 
% $S(m)=-\sum_{\epsilon=\pm 1} \frac{1+\epsilon\,m}{2}\ln\frac{1+\epsilon\,m}{2}$
$S(m)=-\frac{1+\,m}{2}\ln\frac{1+\,m}{2}-\frac{1-\,m}{2}\ln\frac{1-\,m}{2}$ 
is entropy in the mean-field approximation.
The minima of $F_i(m)$ are attractors for the evolution 
(\ref{rangun}). Since $m_\up$ begins to 
increase as $\gamma gt/\hbar$, it relaxes to the smallest positive value
of $m$ where $F_\up(m)$ is minimal. If the temperature is 
not sufficiently low, or
if $g$ is too small, this is the paramagnetic state, $m_\up\to\tanh\beta g$ 
and $m_\down\to-\tanh\beta g$. The result of the measurement cannot then
be registrated robustly, since, after the coupling with $\RS$ is removed,
$\RM$ returns to $m_\up=m_\down=0$. 
However, for sufficiently large $\g$ ($\g>0.08 J$ for $T=0.34J$), 
the paramagnetic state is totally lost
and the relaxation of M leads to a ferromagnetic state with 
magnetization nearly equal to $1$ for $m_\up$, to $-1$ for $m_\down$
($\pm 0.996$ for $T=0.34 J$, $g=0.09 J$).
When the coupling is switched off after relaxation, 
M remains in the vicinity of that state:
the system A=M$+$B is non-ergodic and the memory of its 
triggering by $\RS$ is kept forever.
The duration of the measurement, namely
\BEQ 
\label{taumeas} \tau_{\rm meas}=\frac{\hbar}{\gamma \g}. 
\EEQ
is governed by the establishment of strong correlations of M and S, which
takes place when the magnetization of M 
reaches significant values having the same 
sign as $s_z$. This stage of the process is the slowest of all.
The dimensionless factor $1/\gamma$ expresses that relaxation 
occurs due to coupling with the bath B.
The final stage, after $m_\up$ has become sizeable, is a more 
rapid exponential relaxation with characteristic time $\hbar/\gamma J$, 
 during which the coupling with S is ineffective, and 
which leads to robust registration in a ferromagnetic state.

Altogether, 
the common final state of $\RA$ and $\RS$ after $\tau_{\rm meas}$
has the form (2) with probabilities
$p_\up=r_{\up\up}(0)$, $p_\down=r_{\down\down}(0)$ 
for $i=\up$ or $\down$; the
states $|\up\rangle\langle\up|$ or
$|\down\rangle\langle\down|$
of $\RS$ are correlated with the ferromagnetic states of $\RA$, 
$\CR_\up$ or $\CR_\down$, having positive or negative magnetization,
respectively.

{\it Conclusion}. In spite of the simplicity of the present model,
its exact solution displays all the features, listed in the introduction,
that a quantum measurement should satisfy. We relied on the
statistical interpretation of quantum mechanics, which naturally leads to
describe an ensemble of measurements on an ensemble of systems and 
which should yield all possible outcomes with Born probabilities.
The process follows an elaborate 
scenario involving {\it several time-scales}. At the very beginning,
over the very short time (\ref{taucollaps}), the state of S collapses,
while A is affected only microscopically. This collapse is governed by 
the large value of $N$ and should be contrasted with the standard decoherence 
processes \cite{wh,models}. 
Somewhat later, it is made irreversible, either by means of the 
interaction with the thermal bath (at the decoherence time (\ref{taudecoh})), 
or under the effect of a small randomness in the coupling
of S with M. This stage is invisible, since it corresponds only 
to a disappearance of complicated many-spin
correlations within M; its duration is longer than $\tau_{\rm collapse}$ 
but shorter than the recurrence time $\pi\hbar/(2g)$
which would exist without any dissipation.
Thereafter the statistics of S becomes classical for the two
values $s_z=\pm 1$ and remains unchanged in time. The system S, although
microscopic, is seen by M as an external magnetic field $\pm g$ which is
sufficient to trigger the subsequent evolution of M from its initial
metastable state towards either one of its ferromagnetic states. Because M
is macroscopic, this evolution is slow; its time-scale $\tau_{\rm meas}$
is governed by the bath B (the parameters of which satisfy
$\hbar\Gamma\gg T\gg \gamma J$ and $\hbar\Gamma\gg J>g$). Finally, the
registration becomes permanent owing to the irreversible relaxation of M
into stable equilibrium. This takes place over a time of order
$\hbar/(\gamma J)$ shorter than the time $\hbar/(\gamma g)$
required to leave the metastable state. 
We have focused on parameters for which the model simulates an ideal 
measurement, but the analysis may be extended to cover realistic imperfect 
measurements, for instance if $N$ is not very large or if the observed 
quantity $\hat{s}_z$ is not conserved.

An essential property allowing the process to be used as an ideal
measurement is the {\it macroscopic size of the apparatus}. It is the large
value of $N$ and the quasi-continuity of the phonon spectrum which ensure
the collapse, the breaking of invariance of M which generates its initial 
metastable state and its final two possible stable states,
and the dynamics of A which leads
to registration. The macroscopic nature of $\RA$ thus 
conditions the form (\ref{2}) of the outcoming state, with
its correlations of classical nature between the sign of the magnetization
of M and the final marginal state $|\up\rangle\langle \up|$
or $|\down\rangle\langle \down|$ of S, and with the occurrence of the Born
probabilities $p_\up$ or $p_\down$ for each of these events.
This statistical description of the final state expresses that each particular
experiment yields a well-defined outcome for M, and that the selection
of a given result for M can serve as a preparation of S immediately
after the process in one of its pure states $|\up\rangle$ or $|\down\rangle$.
The {\it emergence of classical probabilities from quantum mechanics} is thus
a macroscopic phenomenon, explained by means of quantum statistical mechanics
and occurring on definite time-scales. It is comparable with macroscopic
irreversibility, which emerges from hamiltonian dynamics 
for many degrees of freedom in the framework
of statistical mechanics. Actually, here also, several mechanisms come into
play, which explain the irreversibility of the measurement process;
but moreover they transform the microscopic non-commutative probabilistic 
description of quantum mechanics into ordinary probabilities for the final 
state.

%\vspace{0.2cm}

A.E. A. and R. B. acknowledge hospitality of the University
of Amsterdam and Th.M. N. of the CEA Saclay.

\end{multicols}
\end{document}